\documentclass[epj]{svjour}
\usepackage{graphicx}

\renewcommand{\bar}[1]{\overline{#1}}
\newcommand{\ie}{{\it i.e.}}
\providecommand{\Journal}[4] {#1 {\bf#2}, #3 (#4)}
\providecommand{\PLB}{Phys. Lett. B} %
\providecommand{\PRL}{Phys. Rev. Lett.} %
\providecommand{\PRD}{Phys. Rev. D}
\providecommand{\NPB}{Nucl. Phys. B} %
\providecommand{\EPJC}{Eur. Phys. J. C} %
\providecommand{\ZPC}{Z. Phys. C} %
\providecommand{\PR}{Phys. Rep.} %
\providecommand{\JHEP}{J. High Energy Phys. } %

\newlength{\colw}
\begin{document}
\bibliographystyle{h-physrev4}
\title{The leading particle effect from light quark fragmentation
in charm hadroproduction}

\author{Puze Gao \inst{1,2} \and Bo-Qiang Ma\inst{3,1,}\thanks{Email address: mabq@phy.pku.edu.cn; corresponding author. }}

%
%
\institute{Department of Physics, Peking University, Beijing 100871,
China  \and Institute of Theoretical Physics, Chinese Academy of
Sciences, Beijing 100080, China \and CCAST (World Laboratory),
P.O.~Box 8730, Beijing 100080, China }

\date{Received: date / Revised version: date}
%
\abstract{
The asymmetry of $D^-$ and $D^+$ meson production in
$\pi^-N$ scattering observed by the E791 experiment is a typical
phenomenon
 known as the leading particle effect in charm hadroproducton. We
 show that the phenomenon can be explained by the effect
 of light quark fragmentation into charmed hadrons (LQF). Meanwhile,
 the size of the LQF effect is estimated from data of the E791 experiment.
 A comparison is made with the estimate of the LQF effect from
 prompt like-sign dimuon rate in neutrino experiments. The influence
 of the LQF effect on the measurement of nucleon strange distribution
 asymmetry from charged current charm production processes is briefly discussed.
\PACS{ 
       {14.60.Pq}{}  \and
       {12.15.Ff}{}
     } 
} 

\titlerunning{The leading particle effect from light quark fragmentation
in charm hadroproduction} \maketitle


\section{Introduction}
\label{intro}

The leading particle effect in charm hadroproduction has been
observed by many
experiments~\cite{WA8293,E76994,E79196,WA9297,otherex}. The main
feature of this effect is the enhancement in the production of the
charmed hadrons that carry the same valence flavor of the incident
hadron in the forward region, \ie, at positive $x_F$, where $x_F$
is the Feynman variable for the produced hadron, $x_F\equiv
P_z^*/P_z^{*max}$, with $P_z^*$ being the momentum along the beam
direction in the c.m. frame of the colliding hadrons. The result
of the E791 experiment~\cite{E79196} is a typical example of such
phenomenon with high statistics. In the E791 experiment, a 500 GeV
$\pi^-(\bar u d)$ beam is incident on a fixed target, and an
obvious excess of $D^-(\bar c d)$ (having the same valence flavor
$d$ with $\pi^-$) over $D^+(c\bar d)$ has been observed. The
asymmetry variable
\begin{eqnarray}
A\equiv {d\sigma(D^-)-d\sigma(D^+)\over
d\sigma(D^-)+d\sigma(D^+)}\label{ASY}
\end{eqnarray}
 vs. $x_F$ and $P_T^2$ is used to
describe the effect, and a clear rise of the asymmetry $A$ with
$x_F$ has been observed.

On the theoretical side, because perturbative QCD (PQCD) predicts
no asymmetry at leading order (LO) and very small asymmetry at
next to leading order (NLO) for charm and anti-charm quark
productions~\cite{Nason89,Frixione94}, the observed asymmetry is
generally attributed to the hadronization processes. The ``beam
drag effect"~\cite{Norrbin98} implemented in the PYTHIA Monte
Carlo explains the asymmetry through the color strings between the
produced charmed quarks and the beam remnants that moving in the
same general direction. The charmed hadron produced through the
decay/collapse of the low mass string may have more energy than
the original charmed quark through the pull of the fast beam
remnants. By adjusting some parameters of the model, they can
reproduce the observed asymmetry. Another explanation is the
intrinsic charm coalescence model~\cite{Vogt96}, in which an
intrinsic charm quark of the projectile combines a  valence quark
of similar rapidity to form the charmed leading particle. However,
this model predicts much smaller asymmetry than what is observed
experimentally. A further work is the heavy quark recombination
mechanism~\cite{Braaten02}, in which a produced heavy (anti)quark
recombines a light parton of similar velocity that participates in
the hard scattering process. The work employs a simple PQCD
$O(\alpha_s^3)$ picture and describes the high $x_F$ and $P_T^2$
distributions of the asymmetry very well. There are still other
works of quark recombination and quark-gluon string
model~\cite{chang03,Arakelyan97}.

Although so much work has been done, we should still explore other
possibilities. As first promoted by Dias de Deus and
Dur\~{a}es~\cite{DDD00} in this field, we investigate the possible
contribution of light quark fragmentation (LQF) into charmed
hadrons. The LQF effect is an old idea originally suggested by
Godbole and Roy~\cite{god8489} to explain the unexpected high rate
of prompt like-sign dimuon production from many neutrino
experiments~\cite{CDHSW,CCFR93LSD}. Although there are deviations
in experiments and later high statistic experiments tend to show a
smaller effect than early experiments, the LQF effect could not be
ruled out experimentally. When the leading particle effect in
charm hadroproduction is observed, Dias de Deus and
Dur\~{a}es~\cite{DDD00} point out that the generally neglected
fragmentation, e.g. $d\rightarrow D^-(\bar c d)$ could be a
possible mechanism to explain the observed asymmetry. When the
NuTeV anomaly~\cite{NTV02} is shown to be hopefully settled by the
nucleon strange
asymmetry~\cite{bm96,oln03,kre04,dm04,alw04,dxm04,wak04}, while on
the other hand, CCFR and NuTeV measurements do not show evidence
for the strangeness asymmetry~\cite{CCFR95,NTmas04}, we point out
that the LQF effect is possible to influence such
measurements~\cite{gm05}. In this paper, we will investigate the
LQF effect in charm hadroproduction, and estimate the size of the
LQF effect from the E791 experiment. In section II, we will
describe the physical mechanism of the LQF effect, and present its
parametrization. A detailed description of our calculation will be
given in section III. The results will be shown and be discussed
in section IV. In section V we will give the summery and
conclusions.

\section{The LQF effect}

In our model of the LQF effect, we differentiate the favored
fragmentation $q\rightarrow \bar D(\bar cq)$ or $\bar q\rightarrow
 D(c\bar q)$, which is non-negligible, from the unfavored
fragmentation, e.g., $q\rightarrow D(c\bar q')$, which could be an
indirect effect from LQF due to charm flavor conservation and can
be neglected, since it is mainly confined near target area, \ie,
in large negative $x_F$, where experiments were intangible. In
$\pi^-N$ scattering, for example, since there are more $d$ quarks
than $\bar d$ quarks in $\pi^-$, more $d$ than $\bar d$ will
participate in the hard scattering, and with the favored
fragmentation of $d\rightarrow D^-(\bar cd)$ and $\bar
d\rightarrow D^+(c\bar d)$, more $D^-$ than $D^+$ will be produced
in the forward region, and this is in accord with what is observed
in experiments.

In this work, the picture of the LQF effect has some differences
with works of other authors. We propose a pick-up mechanism of the
light quark fragmentation. After hard scattering of subprocesses,
the outgoing light quark will experience a period of
hadronization. This period $\triangle T\sim 1/M$ is relatively
longer compared to some processes with large energy transfer
$\triangle t\sim 1/\triangle E$. Then, through large momentum
transfer of the strong interaction, the outgoing light quark can
pick up a charmed (anti-)quark from the nucleon sea, forming a
cluster, which can then decay into a charmed hadron.

Although in the pick-up process, the light quark gives most of its
energy and momentum to the nucleon system, the $Q^2$ of the
interaction can be very small along the initial quark direction,
and thus the process is actually a nonperturbative one and is not
largely suppressed by $\alpha_s(Q^2)$. For a rough estimate, one
can consider a massless outgoing light quark with energy $E_q$ in
the nucleon rest frame. Through the pick-up process, the light
quark turns to a constituent (with constituent mass to be about
0.3GeV) of the produced $D$ meson along its initial direction. It
is easy to get the squared momentum transfer $Q^2\sim 0.5$
GeV$^2$.
The lowest momentum fraction of charmed sea from the nucleon in
this process should be $\xi\sim {m_c/ \Delta E_q}$, and the size
of the LQF effect is roughly proportional to the number of charmed
sea above $\xi$, $n_c\sim \int_\xi^1 c(x)dx$. This will yield a
$E_q$ dependence of the LQF effect, which is nearly a linear rise
with $E_q$. This trend is consistent with the observed prompt
like-sign dimuon rates in experiments\cite{CDHSW,CCFR93LSD}.

From the pick-up process, the outgoing light quark can hadronize
into a charmed hadron with a fraction $z$ of its initial energy
along its initial direction in the nucleon rest frame. The flavor
of the light quark is most probable to be a valence flavor of the
produced hadron.

For a quantitative estimate of the LQF effect, we construct the
parametrization of the light quark fragmentation function
$D_d^{D^-}(z,E_q)$, which describes the fragmentation of light
quark $d$ into $D^-$ meson (including $D^-$ from $D^{*-}$ decay),
with $E_q$ and $z$ being the energy of the light quark and the
energy fraction of the produced $D^-$ meson, $z\equiv E_D/E_q$, in
the nucleon rest frame. We assume charge symmetry
$D_d^{D^-}(z,E_q)=D_{\bar d}^{D^+}(z,E_q)\equiv D_f(z,E_q)$ and
take $D_f(z,E_q)=C(E_q)\cdot P(z)$, where $P(z)$ is a normalized
parametrization and we use the form $P(z)\propto
z^\alpha(1-z)$~\cite{Kart78} with $\alpha$ being a free parameter.
The form of $C(E_q)$ should be an approximate linear rise as we
have mentioned. We take the form $C(E_q)=a\cdot(E_q-E_0)$ when
$E_q>E_0$, where $E_0$ corresponds to an energy threshold for the
LQF effect. We also suppose a suppression for the fragmentation of
low transverse momentum light quarks, which correspond to remote
interactions. Thus we include the suppression factor
$(1-m_H^2/p_{qT}^2)$ with the restriction $p_{qT}^2>m_H^2$, where
$m_H$ is the mass of the produced hadron, and $p_{qT}$ is the
transverse momentum of the light quark. From above, we get the
form of light quark fragmentation function:
\begin{eqnarray}
D_f(z,E_q)=a(E_q-E_0)(1-m_H^2/p_{qT}^2)P(z,\alpha)\;,\label{DLQF}
\end{eqnarray}
where $E_0$, $a$ and $\alpha$ are parameters to describe the size
and the shape of the LQF effect.

\section{Calculations for charm hadroproduction}

For high energy hard scattering of hadrons A and B, the inclusive
cross section for the production of hadron C can be factorized as
\begin{eqnarray}
{E_Cd^3\sigma_{AB\rightarrow CX}\over d^3{\bf P_C}}
=\sum_{abcd}\int dx_1dx_2
f_A^a(x_1,Q^2)f_B^b(x_2,Q^2)\nonumber\\
\times{d\hat{\sigma}_{ab\rightarrow cd}\over d\hat
t}D_c^C(z,Q^2){1\over \pi z}\;,\label{factor}
\end{eqnarray}
where $f_A^a(x_1,Q^2)$ and $f_B^b(x_2,Q^2)$ are the parton
distribution functions, with $x_i$ being the momentum fraction
carried by the parton in the infinite momentum frame (IMF).
${d\hat{\sigma}_{ab\rightarrow cd}\over d\hat t}$ is the
subprocess cross section, with $\hat t$ being the parton level
Mandelstam variable. $D_c^C(z,Q^2)$ is the fragmentation function
of parton c into hadron C, with $z$ being the energy fraction of
parton c carried by the hadron C, $z=E_C/E_c$, in the parton c.m.
frame. $Q^2$ is the factorization scale, which can be taken as the
squared transverse momentum of the subprocess.

Now we consider the process $\pi^-N\rightarrow D^{\pm}X$. Since
charm hadrons is always regarded as the fragmentation product of
charmed quarks, with the LO subprocesses  $gg\rightarrow c\bar c$
and $q\bar q\rightarrow c\bar c$, and charge symmetry assumption
$D_c^{D^+}(z,Q^2)=D_{\bar c}^{D^-}(z,Q^2)$, the produced $D^+$ and
$D^-$ should be symmetric. However, if the light quark
fragmentation into charmed hadrons is non-negligible, additional
contribution from the LQF effect should be considered according to
Eq.~(\ref{factor}). The largest contributions from the LQF effect
are $d+g\rightarrow d +g$, with $d\rightarrow D^-$, and $\bar
d+g\rightarrow \bar d+g$, with $\bar d\rightarrow D^+$. Since more
d than $\bar d$ quarks exist in $\pi^-$ and the nucleon, more d
than $\bar d$ quark will contribute to the process, and more $D^-$
than $D^+$ will be produced, just as experimentally observed.

LO PQCD calculation of charm hadroproduction for fixed target
experiments with factorization formula is a debated area, since
$Q^2$ is generally only a few GeV$^2$ in these energy region, and
higher order corrections could be large. The NLO
calculation~\cite{Frixione94} produces an increasing factor of
about 2 relative to LO calculation, with a similar shape for
differential cross sections. However, since we only aim at a
sketchy estimate of the LQF effect in this work, and since the
asymmetry of Eq.~(\ref{ASY}) that we calculate is a ratio of the
cross sections, whose uncertainties may be cancelled to some
extent, we expect reasonable results within uncertainties from LO
calculation.

The differential cross section for the inclusive production of
hadron C as a function of $x_F$ can be expressed as
\begin{eqnarray}
{d\sigma_{AB\rightarrow CX}\over dx_F} =\sum_{abcd}\int dx_1dx_2dz
f_A^a(x_1,Q^2)f_B^b(x_2,Q^2)\nonumber\\
\times{d\hat{\sigma}_{ab\rightarrow cd}\over d\hat
t}D_c^C(z,Q^2)\left|{dx_F\over d \hat
t}\right|^{-1}\;,\label{sima/XF}
\end{eqnarray}
where $x_F$ is the Feynman variable for hadron C,
$x_F=2P_z^*/\sqrt{S}$, with $P_z^*$ being the momentum along the
incident direction in the c.m. frame of the interacting hadrons A
and B, and $S$ is the squared center of mass energy
$S=(P_A+P_B)^2$.

When the masses of quarks and hadrons are neglected, $P_C=zp_c$
will be hold in any reference frame, and $x_F$ can be expressed as
\begin{eqnarray}
x_F=z{(x_1+x_2)\hat t +x_1\hat s\over \hat s}\;,~\label{XF1}
\end{eqnarray}
where $\hat s=x_1x_2S$ is the squared center of mass energy of the
subprocess. Thus the term $|{dx_F\over d\hat t}|^{-1}$ in
Eq.~(\ref{sima/XF}) is
\begin{eqnarray}
\left|{dx_F\over d\hat t}\right|^{-1}={x_1x_2S\over
z(x_1+x_2)}\;.~\label{t/XF1}
\end{eqnarray}

However, the masses of c quark and $D^\pm$ meson should be
considered in our case, and their effect can be taken as
corrections to Eq.~(\ref{XF1}) and Eq.~(\ref{t/XF1}). The
corrections are different for charmed quark fragmentation and
light quark fragmentation into charmed hadrons.

In the case of charmed quark fragmentation, with consideration of
charmed quark mass $m_c$ and the produced charmed hadron mass
$m_H$,  Eq.~(\ref{XF1}) and Eq.~(\ref{t/XF1}) can be corrected as
\begin{eqnarray}
x_F=z{x_1-x_2\over 2}+ z'{x_1+x_2\over\hat s} (\hat t-m_c^2+{\hat
s\over 2})\;,~\label{XF2}
\end{eqnarray}
and
\begin{eqnarray}
\left|{dx_F\over d\hat t}\right|^{-1}={x_1x_2S\over
z'(x_1+x_2)}\;,~\label{t/XF2}
\end{eqnarray}
where $z'=z(1-\epsilon_H+\epsilon_c)$, with $\epsilon_H={2m_H^2/
(z^2\hat s)}$ and $\epsilon_c={2m_c^2/\hat s}$.

In the case of light quark fragmentation, when the produced
charmed hadron mass $m_H$ is considered, and notice that the
fragmentation function is defined in the nucleon rest frame with
$z$ being the energy fraction of the produced hadron,
Eq.~(\ref{XF1}) and Eq.~(\ref{t/XF1}) can be corrected as
\begin{eqnarray}
x_F=z{(x_1+x_2)\hat t +x_1\hat s\over \hat
s}-2z\epsilon_H^N{(E_A+M)p_{qz}\over S}\;,~\label{XF3}
\end{eqnarray}
and
\begin{eqnarray}
\left|{dx_F\over d\hat t}\right|^{-1}=\left[z{x_1+x_2\over \hat
s}+{z\epsilon_H^Nx_1E_A\over M\hat s}({2p_{qz}\over
E_q}-1)\right]^{-1}\;,~\label{t/XF3}
\end{eqnarray}
where $E_A$ is the incident energy, which is 500 GeV for the
$\pi^-$ beam in the E791 experiment, and $M$ is the mass of the
nucleon as target; $\epsilon_H^N\equiv {m_H^2/ (2z^2E_q^2)}$, and
$E_q$ and $p_{qz}$ are the energy and longitudinal momentum of the
outgoing light quark in the nucleon rest frame,
\begin{eqnarray}
E_q=x_1E_A(1+{\hat t\over\hat s})+{x_2M\over 2}\;,~\label{Eq}
\end{eqnarray}
\begin{eqnarray}
p_{qz}=(x_1E_A+x_2M){\hat t\over\hat s}+x_1E_A+{x_2M\over
2}\;.~\label{pfz}
\end{eqnarray}

Similarly, the differential cross section as a function of $P_T^2$
can be expressed as
\begin{eqnarray}
{d\sigma_{AB\rightarrow CX}\over dP_T^2} =\sum_{abcd}\int
dx_1dx_2dz
f_A^a(x_1,Q^2)f_B^b(x_2,Q^2)\nonumber\\
\times{d\hat{\sigma}_{ab\rightarrow cd}\over d\hat
t}D_c^C(z,Q^2)\left|{dP_T^2\over d \hat
t}\right|^{-1}\;.\label{sima/PT2}
\end{eqnarray}
When the mass effect is neglected, $P_T^2$ and $\left|{dP_T^2\over
d \hat t}\right|^{-1}$ can be expressed as
\begin{eqnarray}
P_T^2=-z^2({\hat t^2\over\hat s}+\hat t)\;,~\label{PT21}
\end{eqnarray}
and
\begin{eqnarray}
\left|{dP_T^2\over d\hat t}\right|^{-1}= {\hat s\over z^2|\hat
s+2\hat t|}\;.~\label{t/PT21}
\end{eqnarray}
In case of charm quark fragmentation, with the inclusion of the
mass effect, Eq.~(\ref{PT21}) and Eq.~(\ref{t/PT21}) can be
corrected as
\begin{eqnarray}
P_T^2=-z'^2[{(\hat t-m_c^2)^2\over\hat s}+\hat t]\;,~\label{PT22}
\end{eqnarray}
and
\begin{eqnarray}
\left|{dP_T^2\over d\hat t}\right|^{-1}= {\hat s\over z'^2|\hat
s+2(\hat t-m_c^2)|}\;,~\label{t/PT22}
\end{eqnarray}
where $z'=z(1-\epsilon_H+\epsilon_c)$, with $\epsilon_H={2m_H^2/
(z^2\hat s)}$ and $\epsilon_c={2m_c^2/\hat s}$. In case of light
quark fragmentation from the nucleon rest frame, with the mass of
the produced hadron $m_H$ being considered,  Eq.~(\ref{PT21}) and
Eq.~(\ref{t/PT21}) can be corrected as
\begin{eqnarray}
P_T^2=-z^2(1-2\epsilon_H^N)({\hat t^2\over\hat s}+\hat
t)\;,~\label{PT23}
\end{eqnarray}
and
\begin{eqnarray}
\left|{dP_T^2\over d\hat t}\right|^{-1}={\hat s\over z^2|\hat
s(1-2\epsilon_H^N)+2\hat t|}\;,~\label{t/PT23}
\end{eqnarray}
where $\epsilon_H^N\equiv {m_H^2/ (2z^2E_q^2)}$, and $E_q$ is
expressed in Eq.~(\ref{Eq}).

In our calculations, $\hat s\geq(2m_H)^2$ and $z\geq
2m_H/\sqrt{\hat s}$ are required in $c\bar c$ production and
fragmentation processes, and $m_H/E_q\leq z\leq
1-(m_{\Lambda_c}-M)/E_q$ is required in light quark fragmentation.
To compare our calculation with the E791 data, restriction
$-0.2<x_F<0.8$, just as in the experiment, is obliged in the
calculation of $P_T^2$ distribution. As to the calculation of
$x_F$ distribution, $P_T^2$ is indirectly restricted by $Q^2\geq
Q_0^2$.

We use CTEQ6L1 parton distributions~\cite{CTEQ6} for the nucleon,
and we employ the LO parametrization forms of Ref.~\cite{Gluck99}
for the parton distributions of the $\pi^-$. Since the E791
experiment uses a target mainly of carbon, we take it as an
isoscalar target for the nucleon parton distributions. We consider
LO subprocesses $gg\rightarrow c\bar c$ and $q\bar q\rightarrow
c\bar c$ for charm quark production, and various LO subprocesses
including $qg\rightarrow qg$, $qq'\rightarrow qq'$, $q\bar
q\rightarrow q\bar q$, $qq\rightarrow qq$, $q\bar q\rightarrow
q'\bar q'$, $gg\rightarrow q\bar q$, for light quark d and $\bar
d$ production. The cross sections $d\hat \sigma\over d\hat t$ for
these subprocesses can be found in
Refs.~\cite{Jones78,Babcock78,Owens78}. We use the LO running
$\alpha_s$ with $\Lambda_{QCD}=215$ MeV for $n_f=4$ and
$\Lambda_{QCD}=165$ MeV for $n_f=5$, as specified in CTEQ6L1. We
set factorization scale $Q^2=p_{cT}^2+m_c^2$ for charm quark
production processes, and $Q^2=p_{qT}^2$ for light quark
production processes, with $m_c=1.5$ GeV and
$Q_0^2=m_H^2\approx3.5$ GeV$^2$, where $p_{qT}$ is the transverse
momentum of quark q. We use Peterson
parametrization~\cite{Peterson83} for charm fragmentation function
$D_c^D(z,\epsilon_P)$, with $\epsilon_P=0.08$ and the
fragmentation fraction 0.23 for $c\rightarrow D^+$ (including
$D^+$ from $D^{*+}$ decay)~\cite{Lellis04,Gladilin99}. For the
light quark fragmentation $d\rightarrow D^-$ and $\bar
d\rightarrow D^+$, we use the parametrization of Eq.~(\ref{DLQF})
and tuning the parameters to describe the observed asymmetry.

\section{Results and discussions}

\begin{figure*}
\includegraphics*[width=\colw]{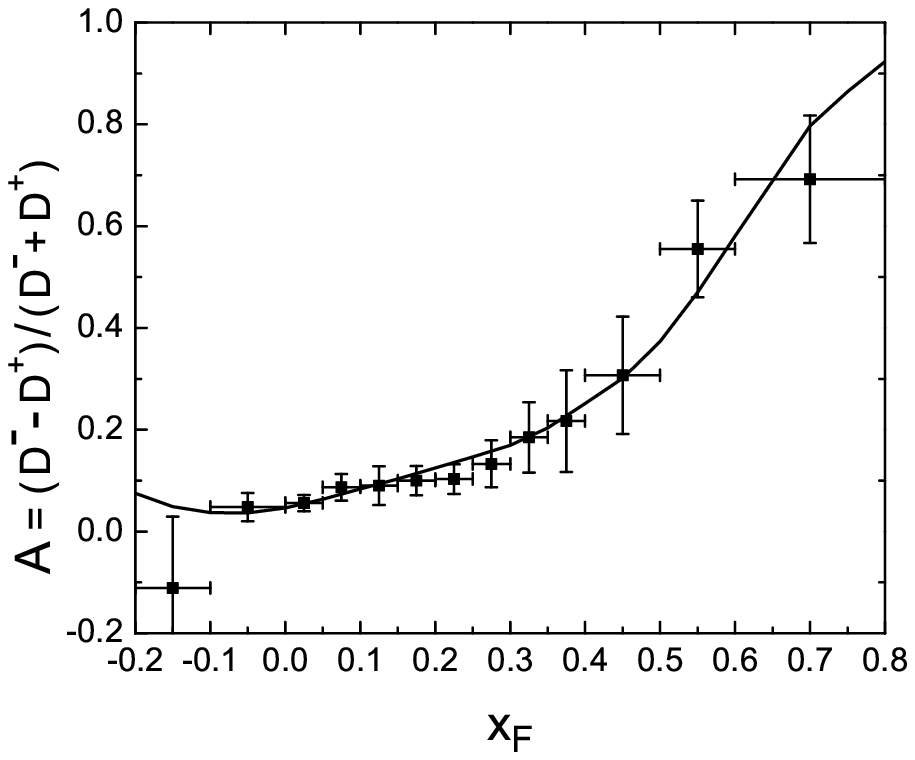}
\includegraphics*[width=\colw]{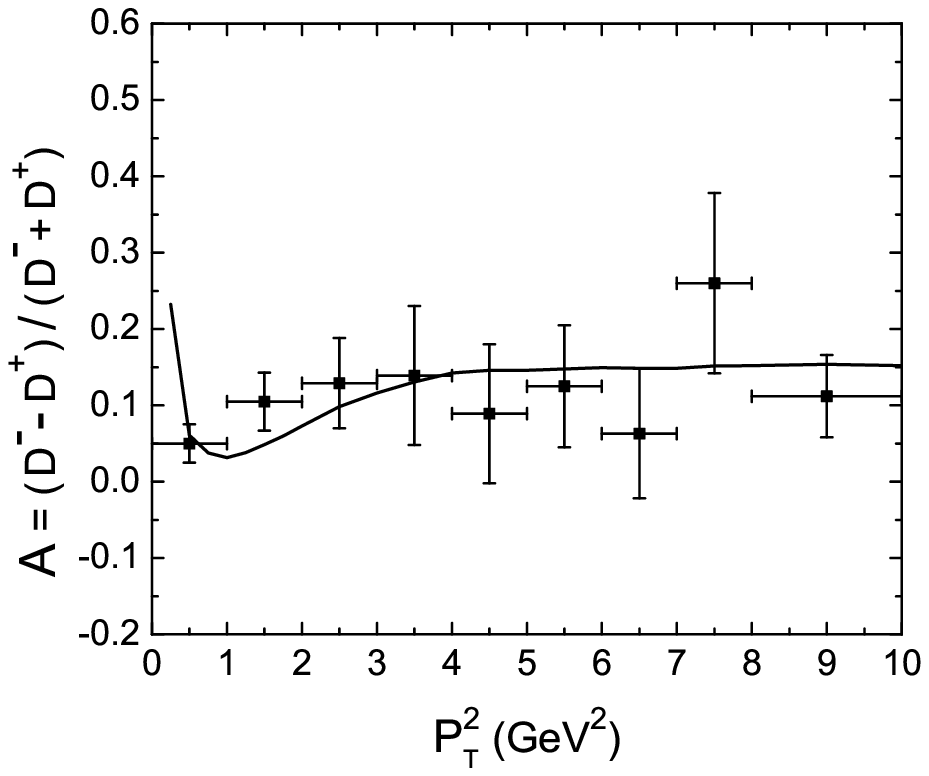}
\caption{ Comparison of the asymmetry variable of Eq.~(\ref{ASY})
from the LQF effect of this work with the E791 data \cite{E79196}.
The solid lines are our calculation results with parameters
$E_0=40$ GeV, $a=2.0\times 10^{-5}$ GeV$^{-1}$, and $\alpha=0.6$.}
\label{fig:ASY}
\end{figure*}

From numerical calculations, we can find good descriptions of the
observed asymmetry from the E791 experiment. Fig.~\ref{fig:ASY}
shows a set of the best fits for both $x_F$ and $P_T^2$
distributions with the following parameters, $E_0=40$ GeV,
$a=2.0\times 10^{-5}$ GeV$^{-1}$, and $\alpha=0.6$. Notice that
our $A(x_F)$ result is consistent with the data of all $x_F$
region, while $A(P_T^2)$ result could not give a good description
of the data below $P_T^2\sim 2$ GeV$^2$. This discrepancy,
however, can be attributed to the unaccounted small random
transverse momentum of the produced hadron relative to the
direction of the light quark when it fragments. When this effect
is considered, the sharp peak of $D^{\pm}$ produced from LQF at
low transverse momentum could become broader, and the asymmetry
curve should get greatly balanced at low $P_T^2$. Meanwhile, the
general features of $A(x_F)$ and $A(P_T^2)~(P_T^2>2)$ do not
change.

$E_0$ is a parameter that describes an energy threshold for light
quark fragmentation into charmed hadrons. There are still
uncertainties in its value, as our result is insensitive to this.
Nevertheless, the overall size of the LQF effect (about
$a(E_q-E_0)$) is quite stable in our fits, when changing $E_0$
from 10 to 80 GeV, to be about $2.4\times10^{-3}$ near $E_q=160$
GeV, which is a typical energy under study. $\alpha$ describes the
shape of the fragmentation function for the LQF effect, and
$\alpha=0.6$ corresponds to a broad peak around $z=0.4$.



In this work, we attribute the leading particle effect for charm
hadroproduction to the LQF effect from the pick-up process of the
produced light quarks. Nevertheless, we do not exclude other
possible contributions that lead to the fragmentation of light
quark into charmed hadrons, which should still be explored by some
dedicated work.

From our LO calculation, the LQF effect could give a good
description of the observed $D^\pm$ asymmetry. On the other hand,
we can get an estimate of the shape and the size of the LQF effect
from the fitting results. When $z$ is integrated out, the size of
LQF effect can be drawn from Eq.~(\ref{DLQF}) to be about
$a(E_q-E_0)$ when $p_{qT}^2\gg m_H^2$, with $E_0=40$ GeV and
$a=2.0\times 10^{-5}$ GeV$^{-1}$, one gets
$D_f(E_q)=2(E_q-40)\times 10^{-5}$. For a typical value $E_q=160$
GeV, one gets $D_f=2.4\times 10^{-3}$, which is very stable in our
fit as we have mentioned above. However, there are still
uncertainties for this fitting result both from LO calculations
and from the parametrization of the light quark fragmentation
function. Since the energy of fixed target experiments is limited,
higher order effects will contribute and the calculated LO cross
sections show a scale dependence. Meanwhile the low transverse
momentum suppression factor $(1-m_H^2/p_{qT}^2)$ in our
parametrization gives an extra uncertainty by a factor of 1/3. As
a sketchy estimate, we set an increasing factor of 2 and a
decreasing factor of 3 as its uncertainties around the central
value of our fit. Thus, for $E_q=160$ GeV, we have an estimate of
$0.8\times10^{-3}\leq D_f\leq 4.8\times 10^{-3}$ \ie
$D_f=(2.4^{+2.4}_{-1.6})\times 10^{-3}$ for the size of the LQF
effect.

Now we compare our result with the previous estimate of the LQF
effect from neutrino induced dimuon production
process~\cite{gm05}. In the previous work, the light quark
fragmentation rate, defined as $D_q\equiv D_q^D+D_q^{D^*}$,
 is different from the definition in this work, which is the
rate of $d\rightarrow D^-$ including $D^-$ from $D^{*-}$ decay,
\ie, $D_f= D_q^D+B^*D_q^{D^*}$, where $D_q^D$ and $D_q^{D^*}$
indicate direct fragmentation into $D$ or $D^*$ meson from light
quarks, and $B^*\approx 1/3$ is the fraction for $D^{*-}$ decay to
$D^-$~\cite{PDG04}. With an estimate for the ratio
$D_q^{D^*}/D_q^D=3/1$ from spin counting, we find that the $D_q$
in previous work should be one half smaller when it is compared to
$D_f$ in this work. $D_q$ can be drawn from Eq.~(13) of
Ref.~\cite{gm05}, $D_q/\bar f_c \approx
0.20~{\sigma_{\mu^-\mu^-}/\sigma_{\mu^-\mu^+}}$, with an estimate
for $\bar f_c=0.86$ and taking
${\sigma_{\mu^-\mu^-}}/{\sigma_{\mu^-\mu^+}}=(3.5\pm1.6)\%$~\cite{CDHSW},
which is the prompt dimuon rate for $100<E_{\mathrm{vis}}<200$ GeV
with cut $p_\mu>6$ GeV. From above, one can get $D_q=(6.0\pm
2.7)\times 10^{-3}$. Now we can compare ${1\over
2}D_q=(3.0\pm1.4)\times 10^{-3}$ with $D_f(E_q=160
~\mathrm{GeV})=(2.4_{-1.6}^{+2.4})\times10^{-3}$, for they have
similar energy range. From above, we find that the size of the LQF
effect from the estimate of this work of charm hadroproduction is
a little smaller but consistent with that of the estimate from
neutrino induced prompt $\mu^-\mu^-$ production process.
Estimation of the LQF effect from $\mu^+\mu^+$ data in the
previous work yields a much larger size, which is not supported by
this work.

In our previous work~\cite{gm05}, we have shown that the LQF
effect could influence the measurement of nucleon strange
asymmetry from charged current charm production processes. When
the LQF effect is large enough, it can balance the effect from
nucleon strange asymmetry, which could explain why CCFR and NuTeV
experiments do not show evidence for the nucleon strange
asymmetry. From the estimate of the LQF effect in this work, the
size of the LQF effect at $E_q=160$ GeV is about
$(2.4_{-1.6}^{+2.4})\times10^{-3}$, which is still twice or more
smaller to compensate the predicted nucleon strange asymmetry that
can explain the NuTeV anomaly~\cite{dm04,alw04,dxm04,wak04}.
However, in CCFR and NuTeV experiments (neutrino energy up to 600
GeV), the energy of $E_q$ can be much larger and the LQF effect is
possible to be greatly enhanced. Thus in CCFR and NuTeV, the LQF
effect is still possible to influence the measurement of nucleon
strange asymmetry to a large extent. Further study of the LQF
effect under various processes will be helpful to further clarify
this effect.

\section{Summery}

In this work, we have attempted to explain the leading particle
effect in charm hadroproduction by the picture of light quark
fragmentation into charmed hadrons (LQF). We can obtain good
descriptions of the observed $D^\pm$ asymmetry from the E791
experiment for both $x_F$ and $P_T^2$ distributions. Although the
uncertainty from our LO PQCD calculation is large, we find a total
size of the order of $10^{-3}$ for the LQF effect at a typical
energy $E_q=160$ GeV of the light quark. This result is consistent
with the estimate of the LQF effect from neutrino induced prompt
$\mu^-\mu^-$ data in our previous work. The influence of the LQF
effect to the measurement of nucleon strange asymmetry in neutrino
induced charged current charm production processes could be
non-negligible. \vspace{0.5cm}
\\
{\bf Acknowledgments: } We thank Prof. K-T. Chao, Prof. C-H.
Chang, C. Meng and F. Huang for helpful discussions. This work is
partially supported by National Natural Science Foundation of
China (Nos.~10421503, 10575003, 10528510), by the Key Grant
Project of Chinese Ministry of Education (No.~305001), and by the
Research Fund for the Doctoral Program of Higher Education
(China).

\end{document}